\documentclass[fleqn,usenatbib]{mnras}
\usepackage{hyperref}

\usepackage{newtxtext,newtxmath}

\usepackage[T1]{fontenc}
\usepackage{ae,aecompl}

\usepackage{graphicx}	
\usepackage{amsmath}	

\usepackage{float}

\usepackage{enumitem}

\usepackage[dvipsnames]{xcolor}

\title[Numerical study of signal suppression and noise]{A numerical study of 21-cm signal suppression and noise increase in direction-dependent calibration of LOFAR data}

\author[M.Mevius et al.]{M. Mevius,$^{1}$\thanks{E-mail: mevius@astron.nl}
F. Mertens,$^{2,3}$,
L. V. E. Koopmans$^{2}$,
A. R. Offringa$^{1}$,
S. Yatawatta$^{1}$,%
\newauthor
M. A. Brentjens$^{1}$,
E. Chapman$^{4}$,
B. Ciardi$^{5}$,
H. Gan$^{2}$,
B. K. Gehlot$^{6}$,
R. Ghara$^{7,8}$, 
A. Ghosh$^{9}$,
\newauthor
S. K. Giri$^{10}$,
I. T. Iliev$^{11}$,
G. Mellema$^{12}$,
V. N. Pandey$^{1}$,
S. Zaroubi$^{7,2,8}$
\newauthor
\newauthor
\\
\\
$^{1}$Astron, Oude Hoogeveensedijk 4, 7991 PD Dwingeloo, The Netherlands \\
$^{2}$Kapteyn Astronomical Institute, University of Groningen, PO Box 800, 9700 AV Groningen, The Netherlands \\
$^{3}$LERMA, Observatoire de Paris, PSL Research University, CNRS, Sorbonne Universit\'e, F-75014 Paris, France
\\
$^{4}$Astrophysics Group, Imperial College London, Blackett Laboratory, Prince Consort Road, London, SW7 2AZ, United Kingdom\\
$^{5}$Max-Planck Institute for Astrophysics, Karl-Schwarzschild-Stra{\ss}e 1, 85748 Garching, Germany\\
$^{6}$School of Earth and Space Exploration, Arizona State University, 781 Terrace Mall, Tempe, AZ 85287, U.S.A.\\
$^{7}$Department of Natural Sciences, The Open University of Israel, 1 University Road, PO Box 808, Ra'anana 4353701, Israel \\
$^{8}$Department of Physics, Technion, Haifa 32000, Israel\\
$^{9}$Department of Physics, Banwarilal Bhalotia College, Asansol, West Bengal, India\\
$^{10}$Institute for Computational Science, University of Zurich, Winterthurerstrasse 190, 8057
Zurich, Switzerland.\\
$^{11}$Astronomy Centre, Department of Physics and Astronomy, Pevensey II Building, University of Sussex, Brighton BN1 9QH, U.K.\\
$^{12}$The Oskar Klein Centre, Department of Astronomy, Stockholm University, AlbaNova, SE-10691 Stockholm, Sweden\\
\vspace{-1em}
}
\date{Accepted XXX. Received YYY; in original form ZZZ}

\pubyear{2020}

\begin{document}
\label{firstpage}
\pagerange{\pageref{firstpage}--\pageref{lastpage}}
\maketitle

\begin{abstract}
We investigate systematic effects in direction dependent gain calibration in the context of the Low-Frequency Array (LOFAR) 21-cm Epoch of Reionization (EoR) experiment. The LOFAR EoR Key Science Project aims to detect the 21-cm signal of neutral hydrogen on interferometric baselines of $50-250 \lambda$. We show that suppression of faint signals can effectively be avoided by calibrating these short baselines using only the longer baselines. However, this approach causes an excess variance on the short baselines due to small gain errors induced by overfitting during calibration. We apply a regularised expectation-maximisation algorithm with consensus optimisation (\textsc{sagecal-co}) to real data with simulated signals to show that overfitting can be largely mitigated by penalising spectrally non-smooth gain solutions during calibration. This reduces the excess power with about a factor 4 in the simulations.  Our results agree with earlier theoretical analysis of this bias-variance trade off and support the gain-calibration approach to the LOFAR 21-cm signal data.
\end{abstract}




\section{Introduction}
Detection of the faint 21-cm signal of neutral hydrogen emitted at high redshifts is one of the hardest radio-astronomical programmes currently pursued. This is in particular so, since the contaminating foreground signals are many orders of magnitude stronger. To extract the 21-cm signal from the data, requires an exquisite understanding of the entire signal chain. 

The Radio Interferometric Measurement Equation (RIME, \citealt{hamaker1996a,smirnov2011a}) is a generalised framework to describe the propagation of the signal of radio sources from the source to the radio interferometer. It includes the full (polarised) signal path, including atmospheric and instrumental signal distortion, in a matrix formalism using Jones matrices. A model of the sky brightness distribution, updated iteratively during gain calibration, predicts the coherence matrix for a given baseline. In its most general form, for a dual polarization instrument, all external effects on the propagation of an electromagnetic wave, are merged into a single time ($t$) and frequency ($f$) dependent complex $2 \times 2$ gain matrix for both elements $i,j$ of an interferometer, which is applied to the sky model coherence matrix to predict the corrupted visibilities $V_{ij}$ of a single baseline: 
\begin{equation}\label{eq:ME}
\mathbf{V}^{\rm pred}_{ij} (t,f) = \sum_{k=1}^{K} \mathbf{J}_i(t,f)\mathbf{C}_{ijk}(t,f)\mathbf{J}_j^{\rm H}(t,f) + \mathbf{N}_{ij},
\end{equation}
where the sum is over $K$ discrete sky model components. $\mathbf{J}_i$ is the $2 \times 2$ complex Jones matrix for antenna $i$, $\mathbf{C}_{ijk}$ is the coherency matrix giving the contribution of each sky model component $k$ to the visibility of baselines $i,j$ and $\mathbf{N}_{ij}$ is a 2x2 complex noise matrix.   

A \emph{solve} step performs a non-linear fit of the predicted visibilities $\mathbf{V}^{\rm pred}_{ij}$ to the measurement to determine the complex gains of the Jones matrices.  For current wide-field low-frequency telescopes, a single Jones matrix per antenna is often insufficient, since atmospheric effects and station beam errors vary over the field of view. These direction-dependent (DD) effects are in general taken into account by dividing the sky model into several patches, with scales over which these effects are thought to vary little, each with its own Jones matrix. This means grouping  the $K$ discrete components in Equation \ref{eq:ME} into $N$ ($N<<K$) directions. In DD calibration the number of free parameters increases linearly with the number of solved directions, enlarging the risk of overfitting. In practice, the sky model will always be incomplete, and during calibration this leads to gain errors whose impact are sometimes difficult to predict. Among these effects are the appearance of \emph{ghost} sources in the images \citep{wijnholds2016}, the increase of noise \citep{PPatil16,Barry16} and the suppression of the signal of the unmodelled sky \citep{Sardarabadi19}. 

In this paper, we study the unwanted signal suppression and noise increase caused by the large number of parameters in direction-dependent calibration through simulation. We investigate this for the specific case, namely the study of the 21-cm signal power spectrum from the Epoch of Reionisation (EoR) with the LOFAR radio telescope. This study serves two goals: first it illustrates the motivation behind the calibration strategy used by \cite{Mertens2020}, thereby verifying the soundness of their results in light of signal suppression. Secondly, it allows us to set a framework for investigating a generic context  radio interferometric gain calibration using many directions.

The EoR refers to the period in the evolution of the Universe in which neutral hydrogen was ionised through irradiation by the first stars and quasars. It is one of the main research areas of modern low-frequency radio telescopes such as GMRT \citep{Paciga13}, LOFAR \citep{vanHaarlem13}, MWA \citep{Morales10} and HERA \citep{DeBoer17}. Although direct detection of the 21-cm signal from the EoR requires instruments with higher sensitivity, the signals are predicted to be detectable through a statistical measurement, typically via its expected signature in a 21-cm power spectrum. This measurement is especially challenging since the power of the expected signal is orders of magnitudes smaller than that of the foreground radiation. It necessitates long integration times (typically thousands of hours) and foreground subtraction with unprecedented accuracy, such that the power of the background noise reaches that of thermal noise. This poses stringent requirements on the level of calibration accuracy. One of the main assumptions of the EoR-signal measurement is that the foregrounds as well as any instrumental or atmospheric effects are smooth in frequency, whereas the signal is not. As will be shown, this assumed smoothness in frequency is important to minimise signal suppression and noise increase in calibration.

The layout of this paper will be as follows. In Section \ref{section:method}, we outline the problem of signal suppression and noise enhancement in the framework of the LOFAR EoR analysis and we present our methods. In Section  \ref{section:calibration}, we summarise the processing steps used in the LOFAR EoR analysis and in our simulation. Section \ref{section:simulation} describes the simulation of the faint 21-cm EoR signals and the subsequent analysis of signal suppression and increased noise power through overfitting. In Section \ref{section:regularisation}, we investigate whether we can reduce both signal suppression and added noise by enforcing smoothness of the gain parameters. Finally, conclusions and recommendations are given in Section \ref{section:conclusion}.

\section{Method}\label{section:method}
We base our analysis on typical data from the Low-Frequency ARray (LOFAR)~\citep{vanHaarlem13}. The LOFAR array consists of 24 core stations, located roughly within a circle with a diameter of 3 km in the Eastern part of the Netherlands, and 14 remote stations with baselines up to $\sim$100 km. A further 14 international stations are not used for the EoR analysis. We use the high-band antenna data of LOFAR, covering frequencies roughly between 110 and 170 MHz.

\cite{Patil17} presented the first upper limit on the 21-cm  signal power spectra from LOFAR. They analysed the data of one of the main LOFAR EoR fields, the North Celectial Pole (NCP).  In their analysis, gain calibration was performed without the baselines that were used for the 21-cm signal measurement (i.e. those shorter than $250\lambda$). We will refer to this as \emph{"applying a baseline cut"} to the data. LOFAR has a sufficient number of longer baselines to determine the (multi-directional) gains per station. After calibration, the gain solutions are applied to all baselines, including the shorter baselines. The reasons for applying a baseline cut are twofold: first, the Galactic diffuse emission that is dominant on those short baselines is not included in the sky model; and secondly, it ensures that no signal suppression occurs in the calibration, since all data that contribute to the 21-cm signal analysis are excluded from the calibration.  However, as was shown by \cite{PPatil16}, overfitting increases the noise level on the baselines that are excluded from the calibration, resulting in an \emph{excess variance} in the 21-cm signal power spectrum. A similar increase in noise is shown by \cite{Barry16}.

Therefore, we have produced a sky model that includes the diffuse emission in the NCP. Here we shall investigate whether the new model allows us to remove the baseline cut to reduce the effect of overfitting. We will also, through simulation, study the effect of removing the baseline cut on signal suppression. \cite{Sardarabadi19} provide a theoretical framework for signal suppression. In this paper we investigate signal suppression via numerical simulations, using the same codes that are used to analyse the LOFAR data.

\subsection{Consensus optimisation}\label{section:consensus}
Various algorithms have been developed that implement the RIME in Equation \ref{eq:ME}, aimed at calibrating radio interferometric instruments. In this work, the software used for calibration is \textsc{sagecal-co} \citep{Yatawatta13, Yatawatta15, Yatawatta16}. \textsc{sagecal-co} makes use of fast distributed systems that include graphics processing units (GPU), and performs consensus optimisation to iteratively force the station-based and direction-(in)dependent gain solutions to approach a spectrally-smooth function. Such smooth gain solutions have previously been shown to be crucial to mitigate signal suppression and noise increase \citep[see][]{PPatil16, Barry16, Ewall17, Sardarabadi19}.  To ensure a spectrally smooth behaviour of the gain solutions, an extra penalty function is added to the  optimisation problem at every iteration. This so called consensus optimisation uses an augmented Lagrangian, with a regularisation parameter to guide the solution to approach the chosen (smooth) regularisation function, which is itself a function of several free parameters. Hence, if chosen wisely, the solutions will exactly match this functional form, reducing overfitting, although this would theoretically take an infinite number of iterations. In practice, a maximum number of iterations and regularisation values need to be chosen carefully to ensure minimal deviations from the smooth regularisation function. \cite{Yatawatta15} gives more details on the implementation. The Lagrangian (eq.~14 in \citealt{Yatawatta15}) to be minimised in \textsc{sagecal-co} is:
\begin{equation}\label{eq:sagecalco}
  {\cal L} = \sum_i g_{fi}(\mathbf{J}_{fi}) + ||\mathbf{Y}_{fi}^H(\mathbf{J}_{fi} - \mathbf{B}_{fi}\mathbf{Z})|| + \frac{\rho}{2}||\mathbf{J}_{fi} - \mathbf{B}_{fi}\mathbf{Z}||^2.
\end{equation}
Here, $g(\mathbf{J})$ is the usual least-squares cost function resulting from the difference between the measured visibilities and the model, given Jones matrices $\mathbf{J}$; $\mathbf{Y}$ is a Lagrange multiplier; $\mathbf{BZ}$ is the smooth function over frequency, with $\mathbf{B}$ a matrix composed of polynomial terms (in frequency) and $\mathbf{Z}$ the fitted parameters; and $\rho$ is a regularisation parameter that determines the level of smoothness in frequency that is enforced on the individual gain solutions during each iteration.  Fitting of the gains is done iteratively. The polynomial coefficients $\mathbf{Z}$ are updated after each iteration. In practice, the number of iterations is limited by computation time. We use a Bernstein polynomial of rank three to regularise the frequency dependence of the gain parameters.  \cite{Yatawatta15} uses simulated data to determine values for $\rho$, quoting typical values of $\rho \in [1,10]$ over a bandwidth of about 60~MHz. The strongest regularisation used in that work is $\rho=50$. As we will show, the regularisation parameter plays a key role in regard to overfitting. We show that increasing $\rho$ to a substantially higher value over smaller bandwidths, greatly improves the result on real data. Possible reasons for this difference could be the additional noise power and the incomplete sky model which were not included in the simulations in \cite{Yatawatta15} and which slow down convergence for real data. In a slightly different simulation in  \cite{Yatawatta16}, an optimal value of $\rho \sim 200$ was found for a fixed number of 50 iterations.

\section{Data Processing and Gain Calibration}\label{section:calibration}

In this section, the LOFAR 21-cm signal data processing steps as followed by \cite{Patil17} and the impact of gain calibration on the 21-cm signal are discussed. Processing of the raw LOFAR visibilities follows several distinct steps, which we will summarise below. For more details we refer to \cite{Patil17} and \cite{Mertens2020}. All analyses are done in the context of the NCP field.

\subsection{Flagging and Calibration}

\begin{description}[align=left]

\item[Step 1 --] The raw visibilities have a 2\,s time and 3.05\,kHz spectral resolution per channel. The data sets are split into sub-bands, each with 64 channels. Each sub-band is cleaned of strong RFI using the package \textsc{aoflagger} \citep{Offringa-2010,Offringa12}. The outer two channels on each side of the sub-band are discarded to avoid spurious signals caused by the poly-phase filter. Visibilities are subsequently averaged to three channels per sub-band, resulting in a 61\,kHz spectral resolution.
\vspace{0.2cm}
\item[Step 2 --] After this initial filtering, flagging and averaging follows the initial gain calibration of the visibilities. A model of the brightest sources for the NCP field, consisting of $\sim$1500 components each with an apparent flux $>35$\,mJy, is used during the initial calibration, and a model of the station beam is applied to the sky model. In  direction independent calibration, only full-Jones station-based complex gain solutions are solved for, in fact solving for the gains in a single effective direction. These gain solutions can then be used to correct the visibilities for station gains and first order atmospheric effects. However, the relatively bright source 3C61.1 is located near the first null of the station beam, causing an apparent flux that is strongly frequency and time dependent. These gain variations can potentially degrade the accuracy of the initial calibration. To mitigate this effect, two separate gain-solutions are solved simultaneously, with one set of gain solutions for 3C61.1 and one for the remainder of the field. The solution intervals over which the gains are assumed constant are 10 seconds and 3 channels (183 kHz). The data is subsequently corrected by applying the latter gain solutions. During initial calibration, only baselines $>30\lambda$ are used to avoid significant diffuse emission affecting the gain solutions. \textsc{sagecal-co} is used over the full 60\,MHz bandwidth, using a third-order Bernstein polynomial to regularise the gain solutions in the frequency direction, but with a relatively low level of regularisation, determined from theoretical considerations in \cite{Yatawatta15}. This level of regularisation allows for fitting of the non-smooth spectral gain variations, such as cable reflections and filter effects, but at the same time reduces overfitting of the thermal noise and sky emission that is not part of the calibration model. 
\vspace{0.2cm}
\item[Step 3 --] Direction-dependent (DD) calibration is done by solving for full-Jones matrices in 122 independent directions, clustered on a sky model consisting of $28,000$ components. The time solution intervals vary between 4 and 20 minutes, depending on the apparent flux in the cluster associated with a direction, and the frequency solution interval is 3 channels (183  kHz). Before DD calibration, the corrected data of the previous step is averaged to a time resolution of 10 s. The frequency resolution remains 61  kHz.

\end{description}

\noindent \textsc{sagecal-co} allows for regularisation of the gain solutions by enforcing the gains to iteratively approach a smooth function in frequency. We make the following choices:
\begin{description}
\item[(i)] During DD calibration, a third-order Bernstein polynomial over the full 60-MHz frequency range is used for spectral regularisation. 
\item [(ii)] The level of regularisation for each iteration is specified by a single parameter per cluster of sources (i.e., a direction). The values of the regularisation parameters $\rho$ are based on the theoretically-estimated required level of regularisation for the centre of the field and are scaled with 0.1 and the relative apparent flux in the clusters.
It is set to $\rho\approx 50$ for the brightest clusters and close to $\rho\approx$1.0 for the faintest clusters. 
\item[(iii)] Each cluster of source components is subtracted from the visibilities after multiplication with their DD gains. This assumes that the gains are constant over the spatial extent of the cluster. 
\end{description}

\noindent In Section \ref{section:regularisation}, it is shown that this level of regularisation still allows for small but significant frequency variations in the visibilities, which are caused by un-modelled sky emission \citep{Barry16}. As the gain solutions have enough freedom to absorb part of the difference between the real sky and the sky model, this causes signal suppression. To avoid this, the baselines used during calibration and signal-extraction are fully separate in all of the current LOFAR 21-cm results \citep{Patil17,Mertens2020}.

\subsection{Baseline selection}\label{subsection:uvcut}

\cite{Patil17} set a lower limit of 
$>250\lambda$  on the baselines used during DD calibration, under the assumption that the unmodelled diffuse emission, dominant on the shorter baselines, could result in signal suppression. In subsequent analysis, a new model, including diffuse galactic emission
modelled at the shortest baselines ($<250 \lambda$) was introduced. In  Section \ref{section:simulation}, we will investigate the effect of the DD calibration scheme with and without the
baseline cut on a simulated 21-cm signal and on the noise power .   

\subsection{Power Spectrum generation}
After calibration and sky model subtraction, the residual visibilities are imaged, using \textsc{wsclean} \citep{Offringa14}. Imaging is performed with settings that are sufficiently accurate for 21-cm power spectra \citep{Offringa19}. Each  frequency channel is imaged individually, resulting in an image cube with dimensions $l,m,f$, with $l,m$ the direction cosines and $f$ the frequency. For the LOFAR-EoR analysis, only the baselines $<250\lambda$ are taken into account. However, for inspection, we also make use of baselines up to $500\lambda$. The most common method to statistically detect the 21-cm signal is to look at the power spectrum of the data in cosmological units \citep{Morales04}. The power spectrum is generated by mapping the $l,m,f$ coordinates to  comoving distances and Fourier transforming to wavenumber ($\mathbf{k}$) space. 
We will present the results as a 2D power spectrum, distinguishing between the modes parallel ($k_{\parallel}$) and perpendicular ($k_{\perp}$) to the line of sight. This enables the possibility to specifically examine baseline (proportional to $k_{\perp}$) and frequency (proportional to $k_{\parallel}$) dependent effects. 

\section{Simulation of faint signals}\label{section:simulation}

In order to test the level of suppression, we add mock 21-cm signals with
 varying amplitudes to simulated data. The simulated data represents the real data after initial calibration (Step 2 in Section \ref{section:calibration}), and includes realistic direction-dependent corruptions. The simulated data are formed by multiplying the 28,000 component model (without diffuse
emission) with the DD gain solutions from real data calibration. Since these corruption were taken from  real data calibration, they were not forced to be spectrally smooth. In reality, we expect the gains, which are mainly the result of instrumental and ionospheric effects, to be spectrally smooth. Therefore, applying strong constraints on smoothness of the gain parameters during calibration, is not expected to introduce additional noise in real data, but it might add some noise in our simulations. This effect of this additional noise power in our simulations will be discussed in section \ref{sec:sim2}, where enforcing smooth gain solutions is studied.   

A realistic level of complex Gaussian
noise is added to the data. The variance is derived from the Stokes V noise in real data after DD calibration. The diffuse foreground emission is simulated
using the model for Stokes I, Q and U that was introduced before for calibration purposes (see Section \ref{subsection:uvcut}). This model consists of shapelets fitted to real data. For practical reasons, the diffuse emission is not corrupted with gain errors. However, in the calibration step it is treated as one independent direction out of many, with its own gain solutions, which are expected to be close to the simulated gains of 1. Therefore, this simplification is not expected to have a large effect on the study presented here. In reality the diffuse emission will be corrupted with direction dependent errors, possibly varying over the extent of the emission. Solving for varying gains over the extended emission is currently not part of the LOFAR EoR calibration scheme, and therefore out of scope of this study. The effect of neglecting gain variations over the extended emission will likely add to the signal suppression in a similar way as an incomplete sky model does. The model that is used for calibration of the simulated data includes all simulated sources and the diffuse emission, but does of course not include the 21-cm signal. This implies that the sky model is complete, i.e., with no emission from unmodelled sources. In \cite{Sardarabadi19} it was shown that an incomplete sky model
during calibration would add to the level of suppression. Therefore, we also performed one test in which we excluded one third of the simulated sources with the lowest flux from the sky model used during calibration. This resulted in no significant difference in the level of suppression with respect to the calibration using the complete sky model. Within our accuracy limits we could therefore not validate that conclusion from \cite{Sardarabadi19}.  Although in reality the sky model will be incomplete, we are confident that the use of the complete sky model in simulations gives a reliable estimate of the minimal level of signal suppression during calibration.

The
simulated 21-cm signal is generated from an image cube covering 115 to 200~MHz, with a frequency resolution of 0.5~MHz and a spatial resolution of 1.17 arcmin, that was originally produced by~\cite{Jelic08} using the {\sc 21cmFast simulation code}~\citep{Mesinger11}. The image data is
linearly interpolated to the LOFAR subband frequencies. Visibilities are
predicted from the image cube using \textsc{wsclean} \citep{Offringa14}. Because a realistic
21-cm signal is orders of magnitude less bright than the noise in a single
night of LOFAR observations, the simulated signal is artificially increased by three orders of magnitude. According to \cite{Sardarabadi19} the level of
suppression is independent of the amplitude of the 21-cm signal, as long as the
signal is faint enough not to significantly alter the gain solutions. To test
the validity of this linearity, we add the signal with 3 different
amplitudes with another factor of $4,7$ and $10$, relative to the original multiplication factor of $1000$ of the simulated 21-cm signal, still one to two orders of magnitude smaller than the simulated foreground signal. Just as the diffuse model, the
simulated 21-cm signal is not corrupted with the \textsc{sagecal} gains. Since the
residual visibilities are the result of subtracting the corrupted sky
model, and not corrected with DD gain solutions, the fact that beam and atmospheric effects are ignored in the 21-cm signal is likely to have a negligible impact on this suppression study. Expanding equation \ref{eq:ME} to specifically include the different components of our model, we obtain:

\begin{equation}\label{eq:MEsimul}
\begin{split}
\mathbf{V}^{\rm pred}_{ij} (t,f) & = \sum_{k=1}^{N} \mathbf{J}_{ik}(t,f)\mathbf{C}_{ijk}(t,f)\mathbf{J}_{jk}^{\rm H}(t,f) \\ &+\mathbf{G}_i(t,f)\sum_{s=1}^{S}\mathbf{DG}_{ijs}(t,f)\mathbf{G}_j^{\rm H}(t,f)\\  &+\mathbf{F}\cdot\mathbf{I}_i(t,f)\mathbf{21cm}_{ij}(t,f)\mathbf{I}_j^{\rm H}(t,f) +\mathbf{N}_{ij}.
\end{split}
\end{equation}
Here $C_{ij}$ refers to the point source foreground model, where each cluster $k$ is the sum over the individual components in the cluster. $DG_{ij}$ is the modelled diffuse galactic emission, composed of $S$ shapelet coefficients. $21cm_{ij}$ is the simulated 21cm signal, multiplied with an artificial constant amplitude $F$.  $G$ and $I$ are set to the identity matrices in our simulation and the $J$ matrices are taken from earlier runs on real data. In the calibration we solve for $J$ and $G$.

 \subsection{Suppression measurement}\label{section:suppression}

\begin{figure*}
	\includegraphics[width=0.45\textwidth]{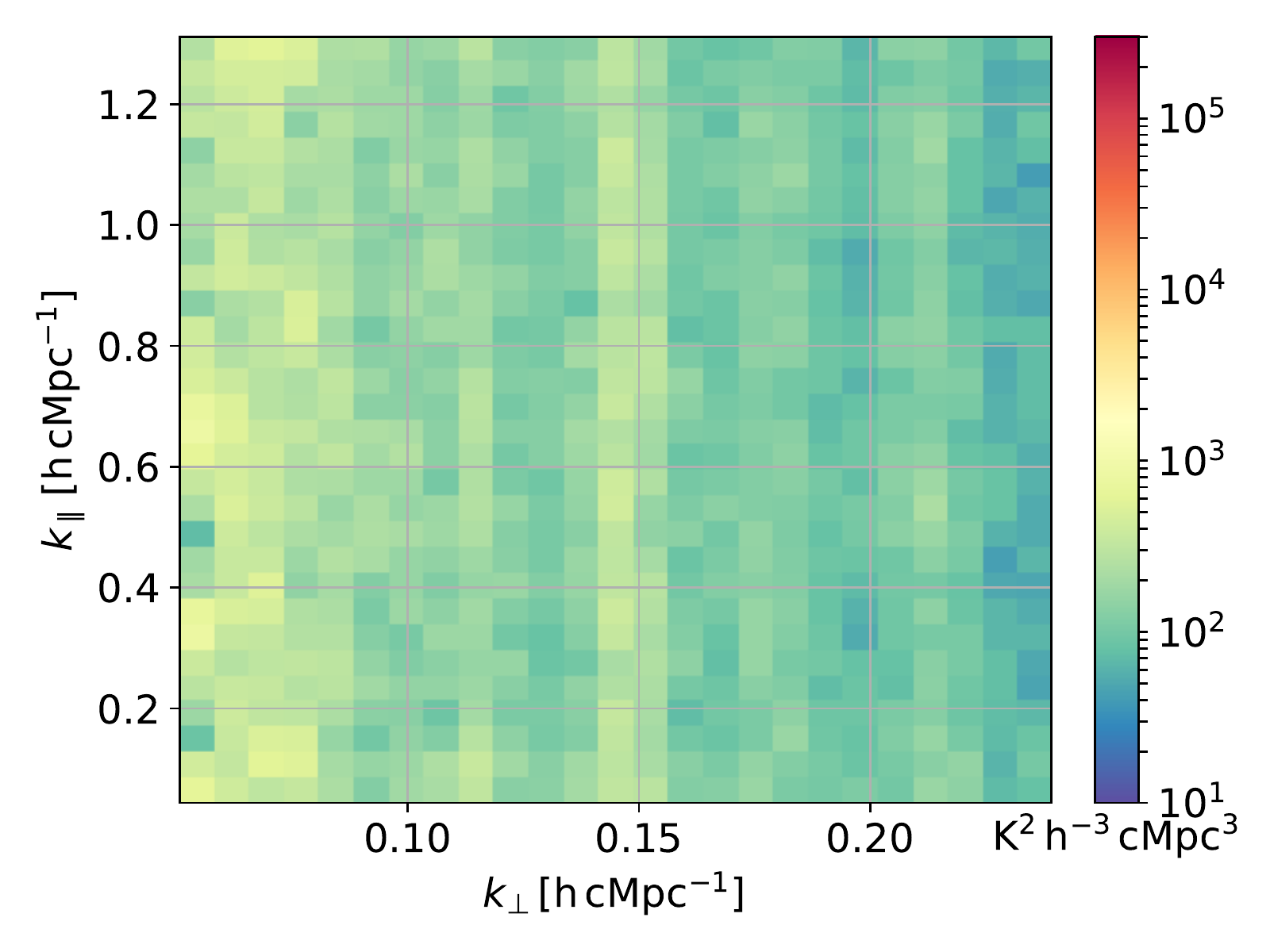}
	\includegraphics[width=0.45\textwidth]{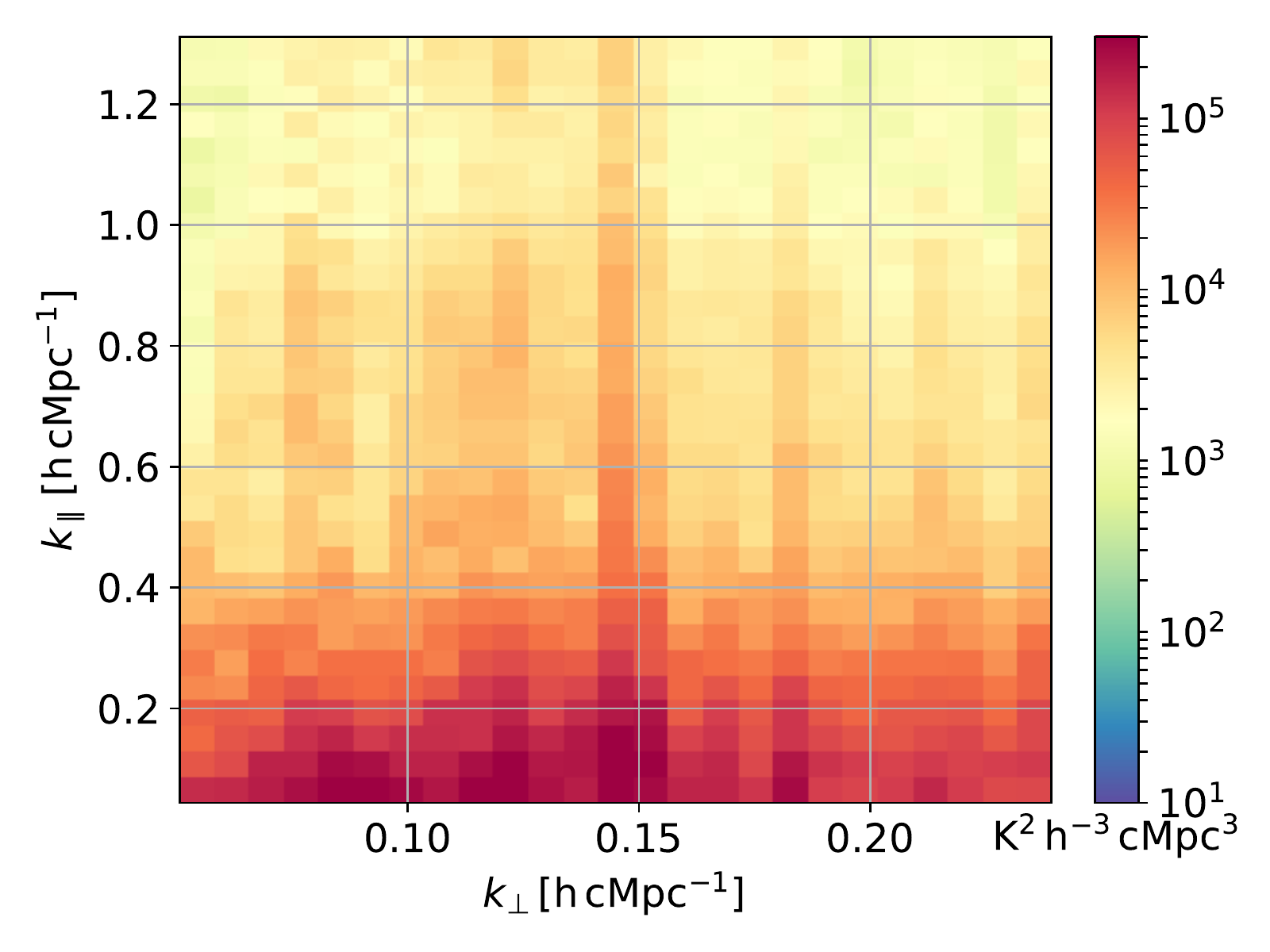}\\
	\includegraphics[width=0.45\textwidth]{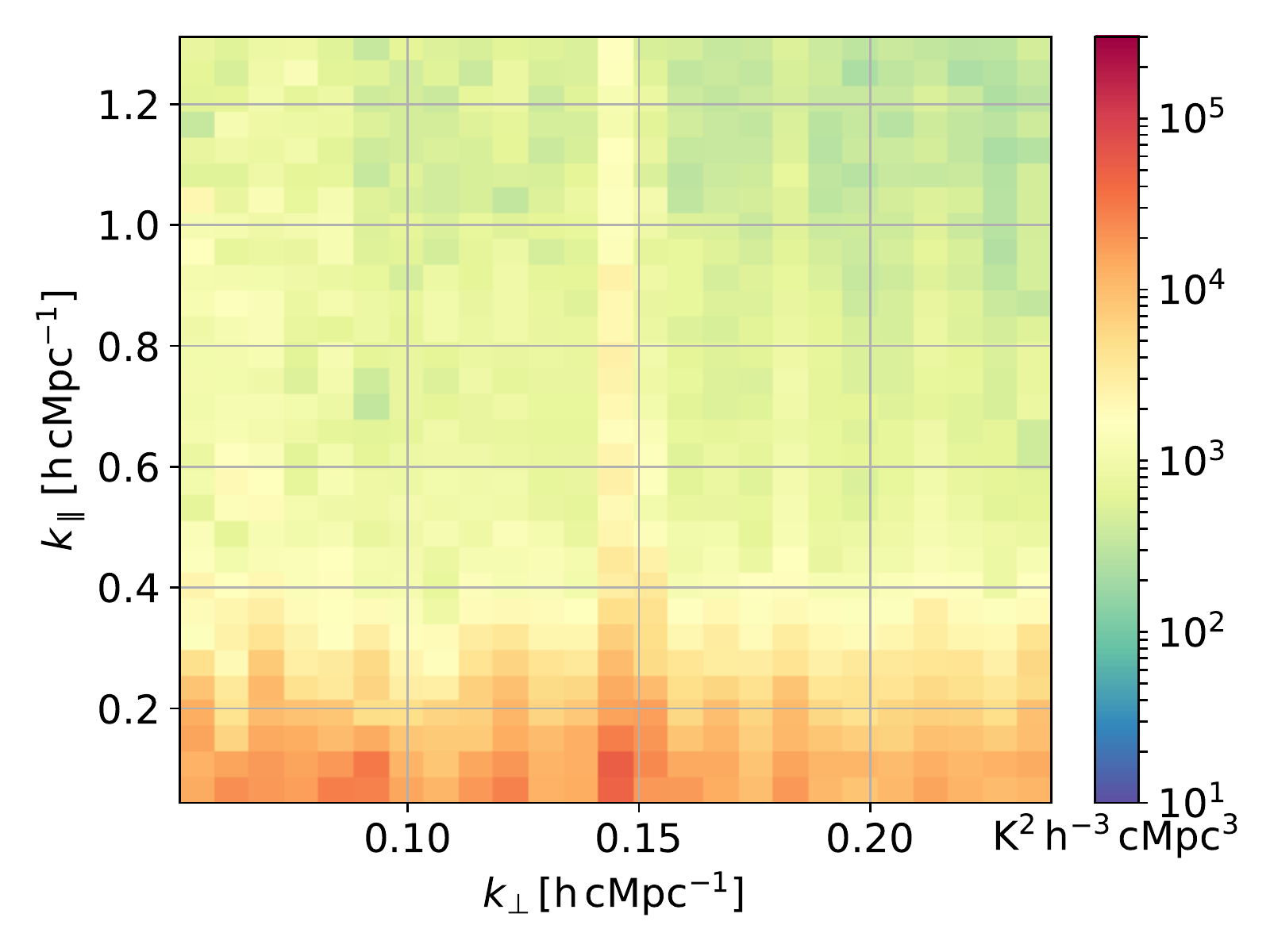}
	\includegraphics[width=0.45\textwidth]{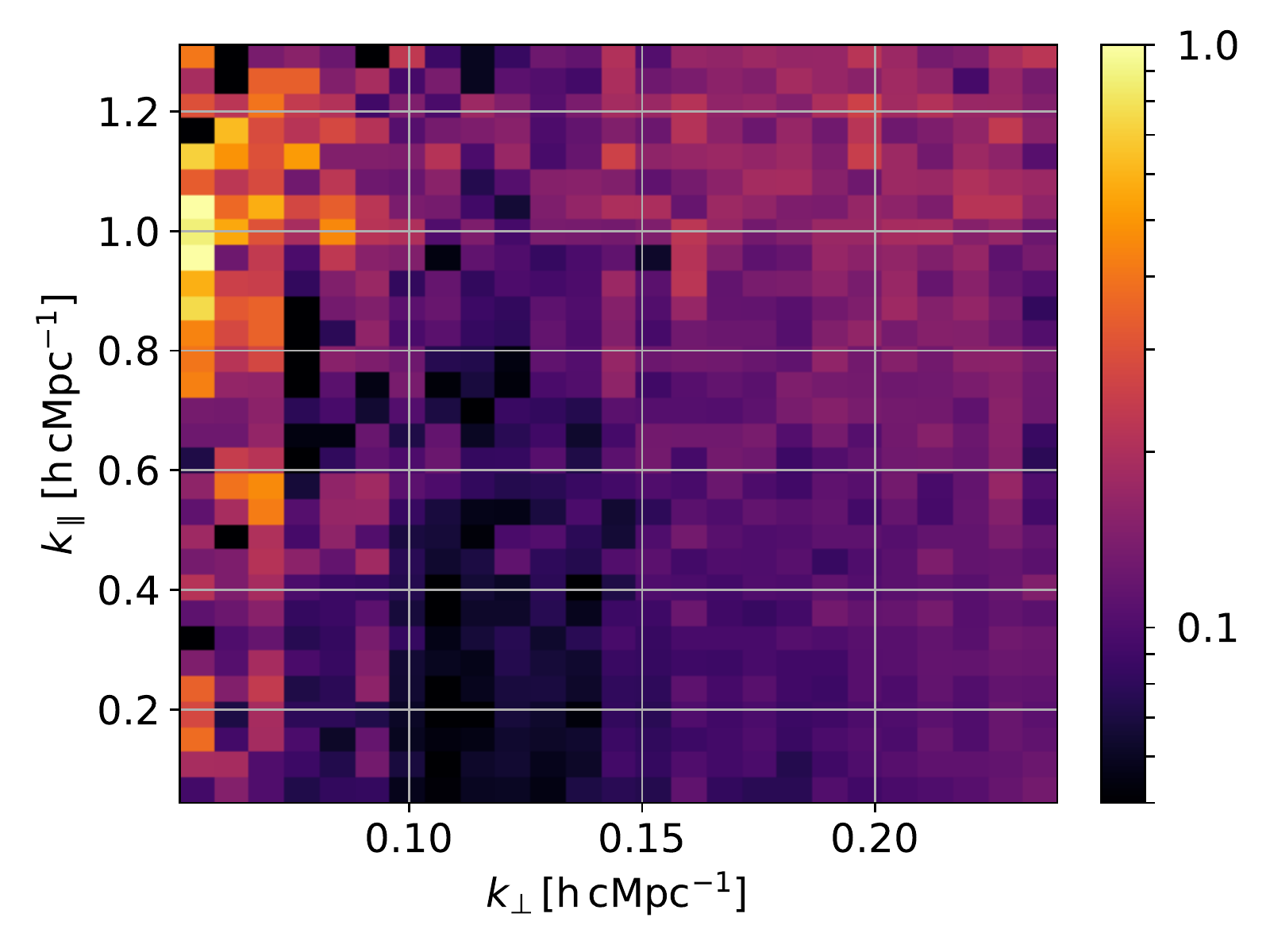}
    \caption{Top left: Power spectrum of the solver noise. Top right: simulated 21-cm signal (relative amplitude 10, see text for explanation). Bottom left: Remaining 21-cm signal after DD calibration. Bottom right: Ratio of 21-cm signal after and before calibration. DD calibration was performed without the 250 $\lambda$ cut on baseline length }
    \label{fig:powerspectra1}
\end{figure*}

We run \textsc{sagecal-co} \citep{Yatawatta16} with standard settings, but without the 250 $\lambda$ baseline cut, on the simulated
data with and without an added 21-cm signal. The computationally intensive DD calibration step for a  typical observation ($\sim$14 hours duration, ~60MHz bandwidth) takes about 72 hours when processed using  half of the dedicated high performance GPU cluster \citep{Pandey2020}. This limits the possible number of independent simulation runs that can be executed in a reasonable time.  We make sure that, by construction, a solution exists. It will, however, not be
recovered precisely because noise has been added and the calibration is a non-linear process, which causes the gain solutions to be inexact. The latter point causes additional noise in the solutions, which we will refer to as the \emph{solver noise}. Since \textsc{sagecal-co} uses
random parameter initialisation, we can estimate the power of the solver noise
by running \textsc{sagecal-co} twice on the same simulated data and calculating the
power spectrum of their difference. We assume the power of the solver noise to be consistent between runs. The power spectrum of the solver noise calculated this way is shown in the top left panel of Fig.~\ref{fig:powerspectra1}.
The top right panel of Fig. \ref{fig:powerspectra1} shows the power spectrum of the simulated 21-cm signal before DD
calibration. The bottom left panel of Fig. \ref{fig:powerspectra1} shows the power spectrum of the simulated 21-cm signal after calibration, and the
bottom right spectrum shows the ratio of the power spectra of the simulated signal after and before DD calibration. The solver noise is one to two
orders of magnitude smaller than the added signal, but it does have an
impact on the ratio in the bottom left plot, albeit minor. In general, due to the solver
noise, the power spectrum after \textsc{sagecal} will be enhanced and therefore
the level of suppression will be underestimated. We compensate for this by
subtracting the power of the solver noise from the signal
power per bin.  As can be seen from Fig.~\ref{fig:powerspectra1}, the suppression of the
signal is large, roughly one order of magnitude. Note that in this suppression measurement all baselines are included in DD calibration and the level of regularisation is relatively low, in contrast to the analyses presented in \cite{Patil17} and \cite{Mertens2020}.  The results with three different levels of input 21-cm signal are summarised in the top three rows of Table \ref{tab:suppression}. The level of suppression is measured by averaging the power over all 750 $k$ bins and taking the ratio with the same average of the input signal. They show that the level of suppression is more or less independent of the signal amplitude, as predicted by \cite{Sardarabadi19}.  There is a small correlation between signal strength and its measured suppression, leading to slightly more suppression on smaller input signals,  but it is negligible compared to the overall level of suppression, and our current measurement is therefore a realistic indication of the level of suppression expected for signals with much smaller amplitudes, such as a true 21-cm signal.

\subsection{Overfitting}\label{section:leverage}

\begin{figure*}
	\includegraphics[width=0.45\textwidth]{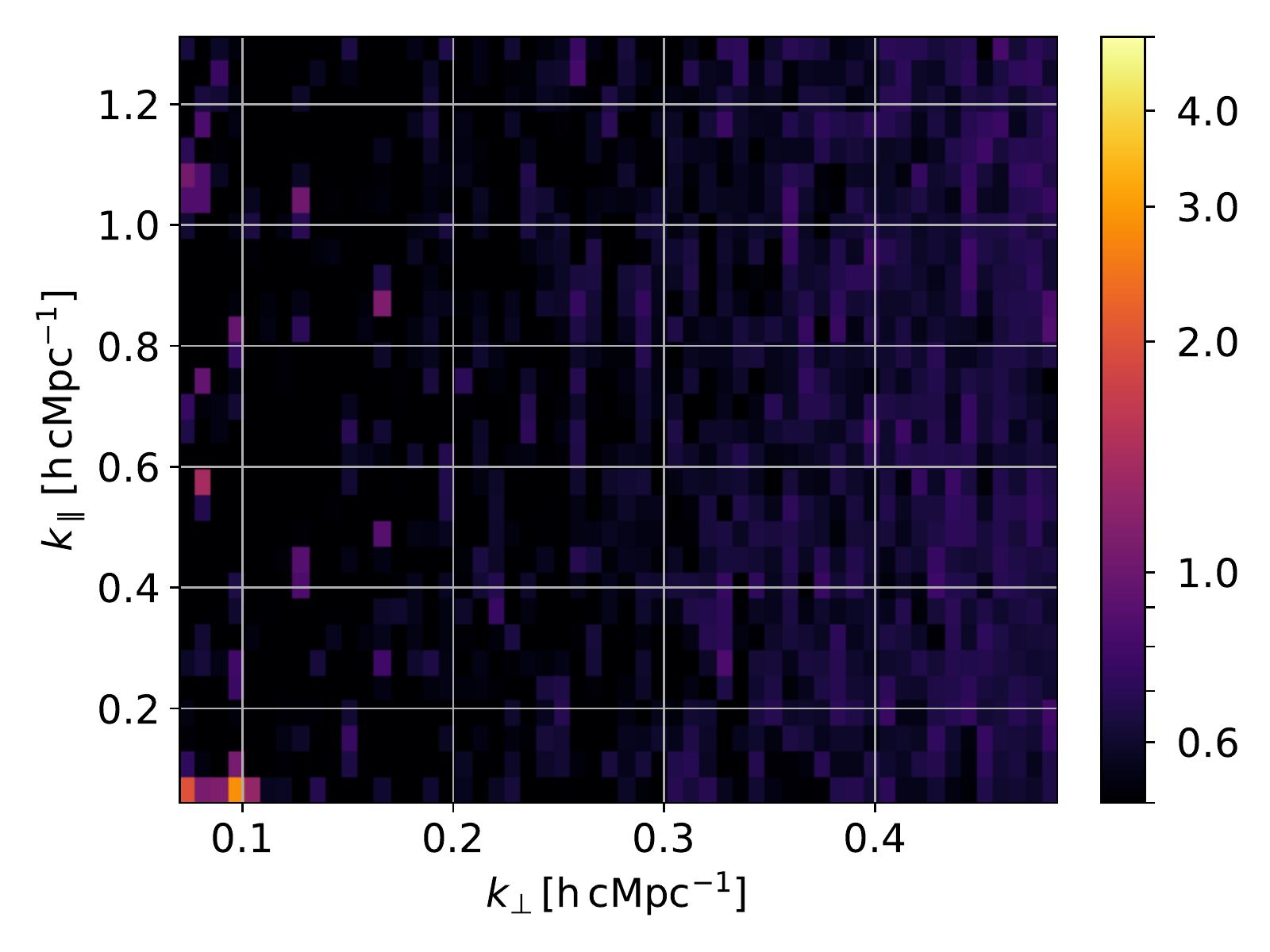}
	\includegraphics[width=0.45\textwidth]{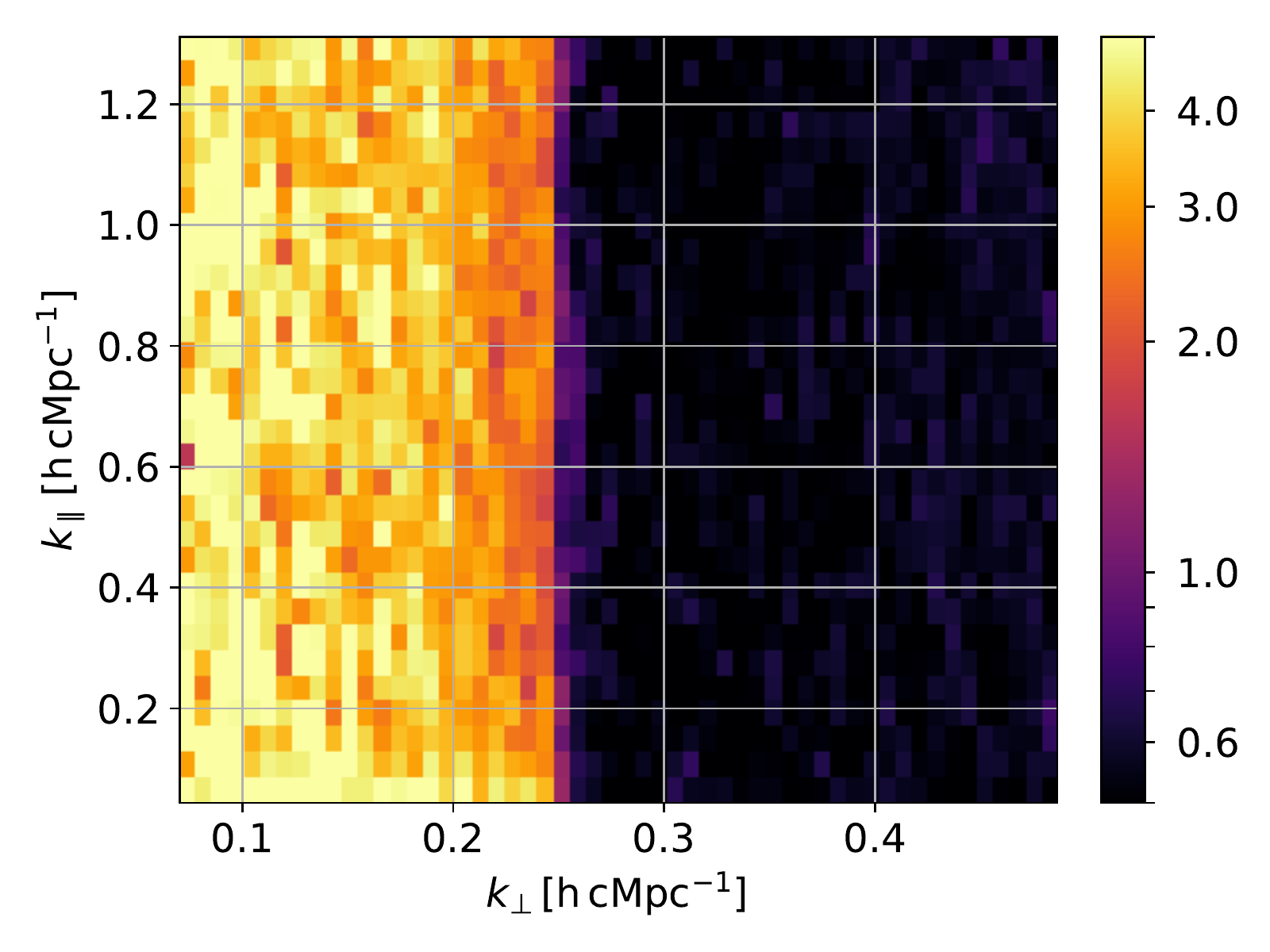}\\
    \caption{Ratio of the Stokes V power spectra after DD calibration and the input noise. Left: All baselines used during calibration. Right: calibrated with baselines larger than $250\lambda$.}
    \label{fig:leverage1}
\end{figure*}

From Section \ref{section:suppression}, it is clear that not using a $250\lambda$ baseline cut leads to unacceptable
levels of signal suppression, even when diffuse emission
is included in the model. Here we test the effect on the signal and noise
when a baseline cut is applied. As discussed in \cite{PPatil16} and
\cite{Yatawatta15}, excluding
baselines from the calibration will increase the variance on these baselines. The significance of
this effect is illustrated in Fig.~\ref{fig:leverage1}, where the Stokes-V power spectrum ratios after and before \textsc{sagecal} are shown over a larger range of scales (between $50$ and $500 \lambda$). For LOFAR, the Stokes-V correlations consist almost only of system noise, and Stokes V allows us therefore to analyse the effect of calibration on a noise-like signal. In the left panel of Fig.~\ref{fig:leverage1}, we show the power ratio of the settings as described in Section \ref{section:simulation}, using a diffuse model and all baselines included during calibration
with \textsc{sagecal-co}. We see a noise power ratio that is
close to one. However, when applying the $250\lambda$ baseline cut, as shown in the right panel of Fig.~\ref{fig:leverage1}, a clear increase of the noise up to a factor $\sim$five
in power is observed on the largest scales. The turn up at $250\lambda$,
corresponding to $k_{\perp}$ of 0.25, is clearly visible. This effect is a
sign of overfitting of the data. Part of the noise is absorbed into the
gain solutions and therefore transferred to the shortest baselines. Note that in our simulations the sky model is complete, so in this case only the noise adds to overfitting of the data \citep{PPatil16,Barry16}. In the following section,
we will investigate whether adding more spectral constraints to the gain solutions by
increasing the regularisation will reduce the level of signal
suppression and/or the overfitting effect.

\begin{figure*}
	\includegraphics[width=0.33\textwidth]{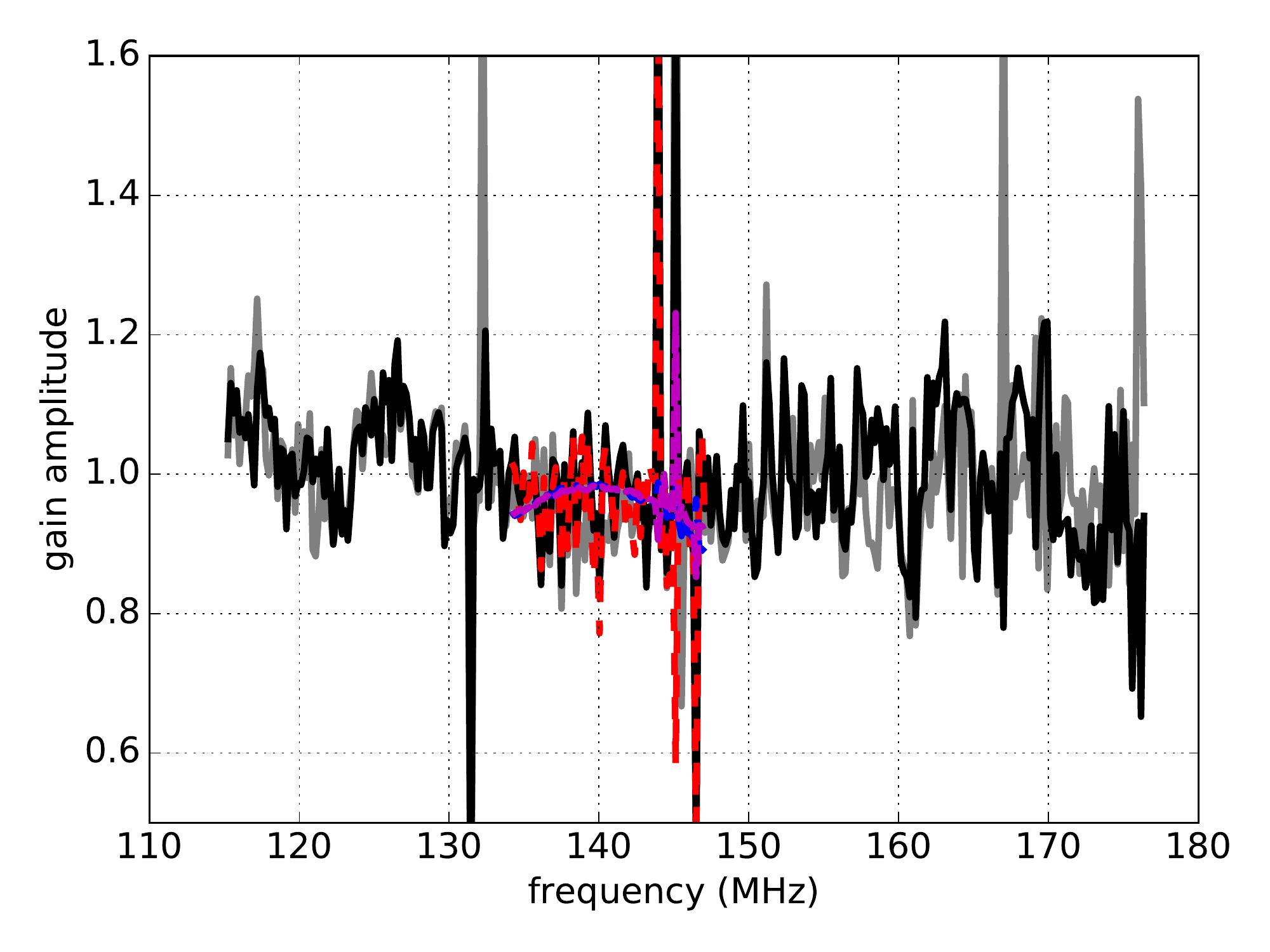}
	\includegraphics[width=0.33\textwidth]{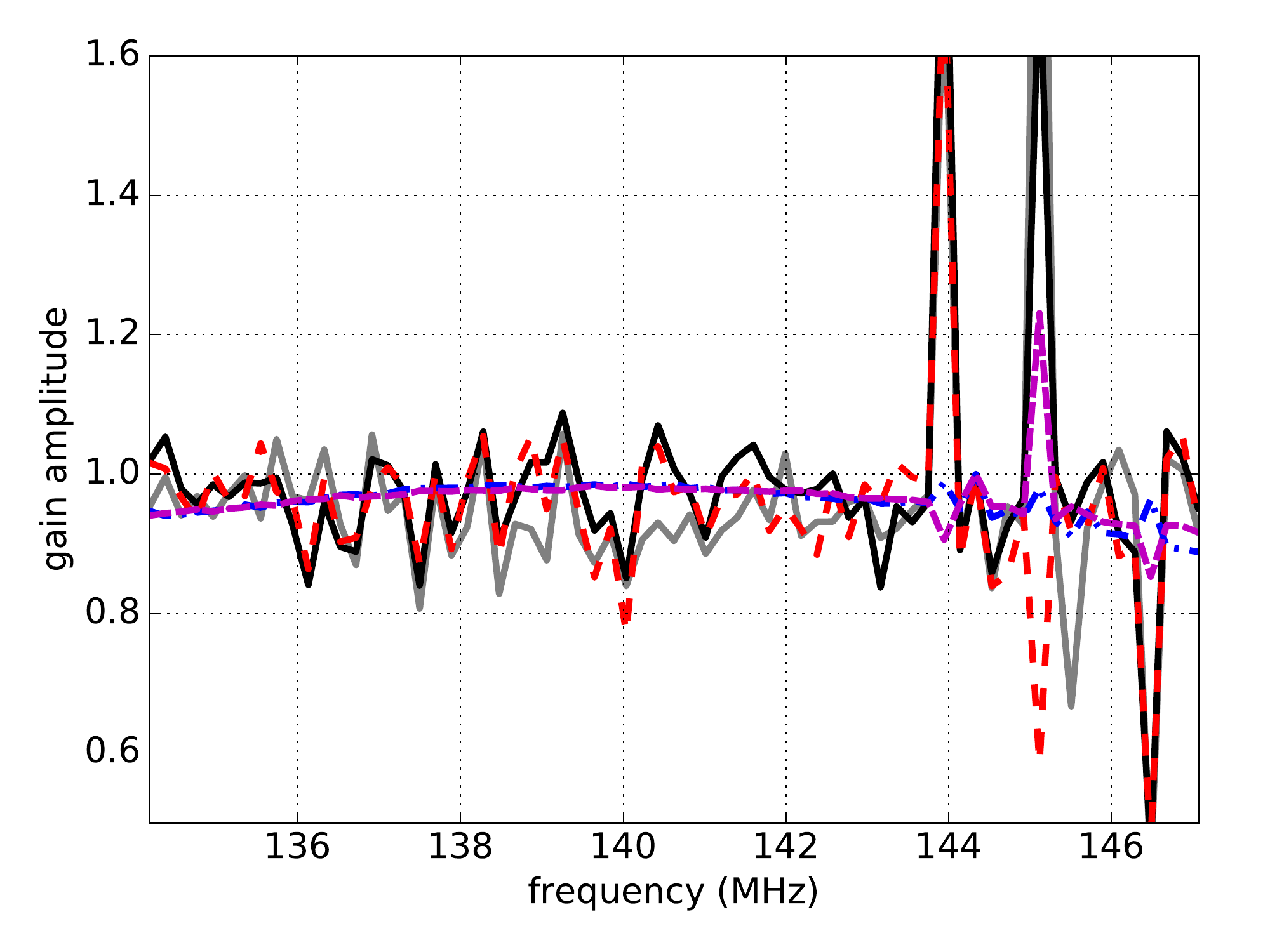}
	\includegraphics[width=0.33\textwidth]{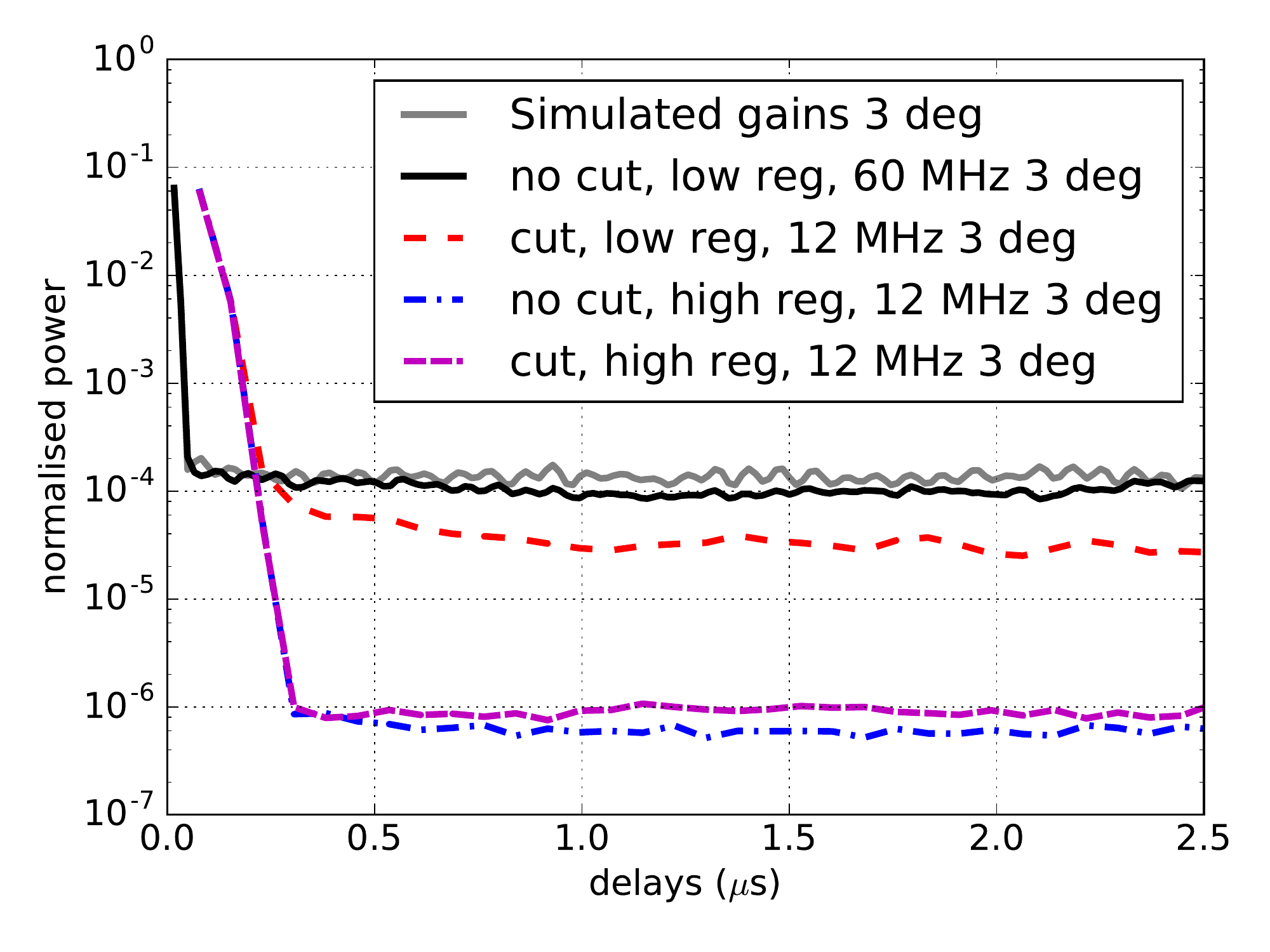}
    \caption{left: Input amplitudes and amplitude solutions for a single timeslot, station and direction ($\approx 3$ degrees away from the field center), using the four different calibration schemes described in \ref{section:simulation} and table \ref{tab:suppression}. Even though a spectral Bernstein polynomial constraint was applied during calibration, the solutions still fluctuate in frequency when the regularisation parameter \textsc{sagecal-co} is low. The large outliers correspond to subbands that had a lot of RFI (in real data) and therefore many of the data were flagged.  The simulated gains were taken from real data calibration with low regularisation and therefore have spectral structure. Middle: zoom of the left plot on the 12 MHz used in the latter 3 calibration schemes. A clear improvement in the smoothness of the solutions is observed when high regularisation is applied. Right: time averaged delay spectrum of the complex gain solutions, same cluster and station. The reduction in noise power with higher regularisation is obvious. Note the different binning  due to the 5 times larger bandwidth of the first two lines.}
    \label{fig:solutionsDD_low}
\end{figure*}

\section{Spectral regularisation}\label{section:regularisation}
The calibration scheme discussed in Section \ref{section:simulation} has also been
applied on real data. As expected, removing the 250 $\lambda$ cut considerably reduces the
noise in the Stokes I and V power spectra, compared to \cite{Patil17}. However, it also has a major effect on the 21-cm signal
itself. The solid black line in Figure \ref{fig:solutionsDD_low} left shows the DD gain variation for one timeslot and one
of the 122 directions. From this figure it is apparent that the final DD
solutions are not spectrally smooth on MHz scales. From the physical processes involved (i.e. the ionosphere and the instrument response), one would expect the true gains to be spectrally smooth up to scales of 10~MHz or larger. The overfitting discussed in the
previous section manifests itself in the irregularity of the solutions on much smaller frequency scales. As
discussed by \cite{Sardarabadi19}, smoother solutions may mitigate the effect of signal
suppression and the excess variance. We can enforce smoothness by increasing either
the number of iterations or the value of the regularisation parameter in \textsc{sagecal-co}. A third order Bernstein polynomial
might also be too constraining to describe the expected frequency variation over the
full 60-MHz bandwidth. Therefore, we reduce the bandwidth to 12 MHz, covering only 134 to 146 MHz, corresponding to the $z=9.6 - 8.7$ redshift bin used by both
\cite{Patil17} and \cite{Mertens2020}. By decreasing the bandwidth, a third-order polynomial provides enough freedom to capture the shape of the bandpass.  Without any regularisation, the number of gain parameters is reduced by a factor 5 due to the limited bandwidth, but so is of course the number of input visibilites. Full regularisation, forcing the solutions to lie on the Bernstein polynomial, would reduce the number of degrees of freedom by a factor 20, since there are 62 subbands within this 12 MHz that have a individual gain solutions. Since in our experiments full convergence is not always met, the number of degrees of freedom will only be partly be reduced by regularisation. A more formal method to derive the remaining number of degrees of freedom is given in \cite{Yatawatta19}.

For a few intervals of real data, we investigate the effect
of the number of iterations and the regularisation
parameter on the smoothness of the solutions. From theory \citep{Sardarabadi19},
one expects that the solutions will be constrained to a smooth
curve once convergence is reached.  Selecting a regularisation parameter that is too low
may increase the number of iterations needed to reach convergence, whereas a high
regularisation parameter may force the solutions to a wrong minimum every iteration,
thereby slowing convergence. We find that with the default \textsc{sagecal-co} setting for the regularisation parameter $\rho$, with values between 1 and 50, depending on the apparent cluster brightness, convergence to smooth gain solutions is not
reached even after 400 iterations. Because the number of
iterations is in practice constrained by compute power limitations, we decided to use 40
iterations and find the optimal regularisation parameters for that
setting. 

\begin{figure}
	\includegraphics[width=\columnwidth]{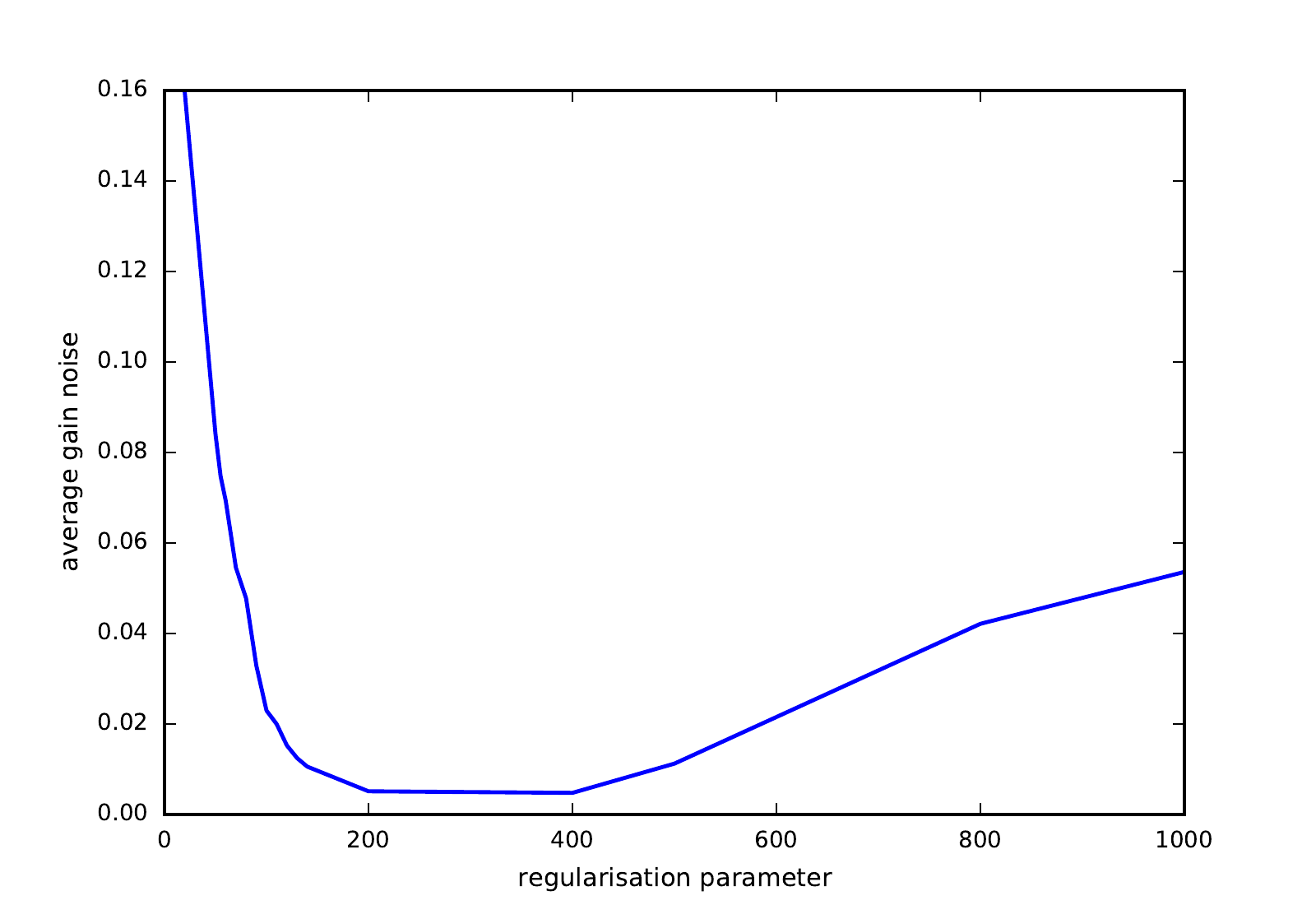}
    \caption{Demonstration of the effect of the regularisation parameter on the average gain variance, using the \textsc{sagecal-co} solutions from a single cluster (real data).}
    \label{fig:gainvariance}
\end{figure}

\begin{figure*}
	\includegraphics[width=0.8\textwidth]{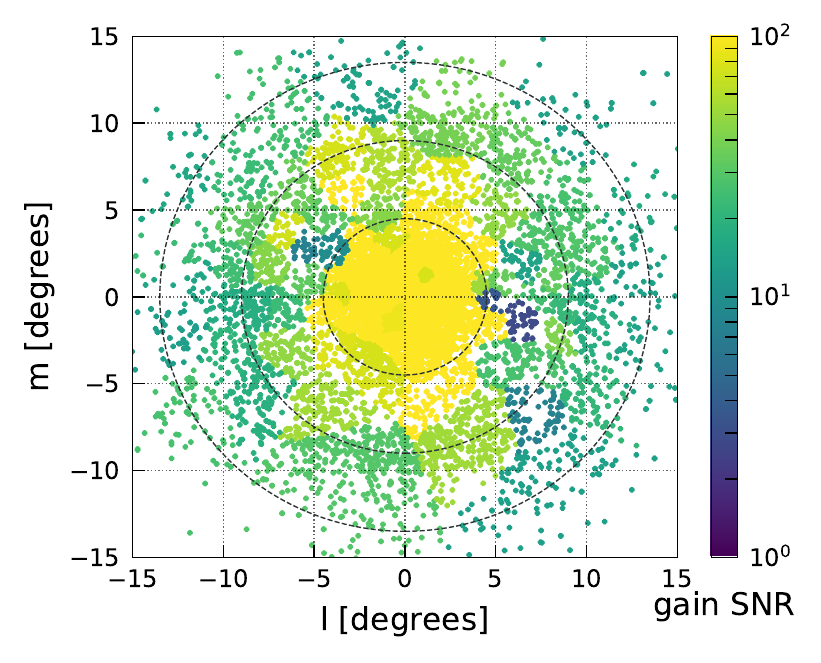}
    \caption{S/N-ratio of the subband differenced solutions for all $\sim 28.000$ sky components in real data. Clusters of components can be recognised since they share the same gain solutions. In total there are 120 clusters in the image. The S/N ratio is similar for most clusters, but showing a slight gradient to lower S/N values away from the phase centre. Some outliers near the first null of the beam can be observed.}
    \label{fig:clustervariance}
\end{figure*}

A good measure of the smoothness of the gain solutions can be
obtained by examining the differences of the gains of adjacent subbands. In 
Fig.~\ref{fig:gainvariance}, we present the variance of the subband differenced gain
solutions for a typical cluster as a function of regularisation parameter. To obtain this plot we used real data. A clear minimum in differential gain variance is obtained near the regularisation
value of $\rho = 200$, which is 200 times  higher than the original regularisation parameter
for this specific cluster. Similar optimal values were found in \cite{Yatawatta16}. In the same way, we find that for most of the 122 directions, the optimal regularisation parameter is tens to hundreds times higher than the initial regularisation. Only a few clusters that are close to the first null of the beam
do not show such a clear minimum of differential gain variance. A possible explanation could be
that a beam model is not applied to the sky model during DD calibration. This results in stronger
frequency-dependent beam variations near the first null, causing spatial gain variations over the extent of the cluster not constant enough to be described
by a single Jones matrix. Including the beam model is one of the envisioned improvements
to our current calibration. The signal-to-noise (S/N) ratio of the gains using real data, defined as the absolute gain divided by the subband differenced gain variance, using the optimised regularisation parameters, 
is shown in Fig.~\ref{fig:clustervariance}. Every point in this figure is a
single component of the 28000 sky model components.
Clearly, clusters that share the same direction-dependent gain solution show the same S/N-ratio.  The S/N ratio of the clusters shows a slight gradient to lower S/N values away from the phase centre, possibly related to the lower apparent flux at those positions. Some outliers with lower S/N values are visible near the first null of the beam. 

\begin{table*}
	\centering
	\caption{Remaining simulated 21-cm signal power after DD calibration with different settings.
          The calibration was performed with and without applying the 250 $\lambda$ cut and with a different regularisation parameter and bandwidth.}
	\label{tab:suppression}
	\begin{tabular}{lccc} 
		\hline
		 DD calibration scheme & Relative signal amplitude & Recovered power ratio&
                Solver noise fraction\\
		\hline
		no cut, low reg, 60 MHz & 4 & $8.3\pm 0.2\%$ & $15\%$\\
		no cut, low reg, 60 MHz & 7 & $8.7\pm 0.2\%$ & $6\%$\\
		no cut, low reg, 60 MHz & 10 & $9.1\pm 0.2\%$ & $4\%$\\
		\hline
		cut, low reg, 12 MHz & 10 & $103\pm 3\%$ & $16\%$\\
		no cut, high reg, 12 MHz & 10 & $65\pm 2\%$ & $5\%$\\
		cut, high reg, 12 MHz & 10 & $100\pm 2\%$ & $6\%$\\
		\hline
	\end{tabular}
\end{table*}

\subsection{Simulations}\label{sec:sim2}

\begin{figure*}
	\includegraphics[width=0.45\textwidth]{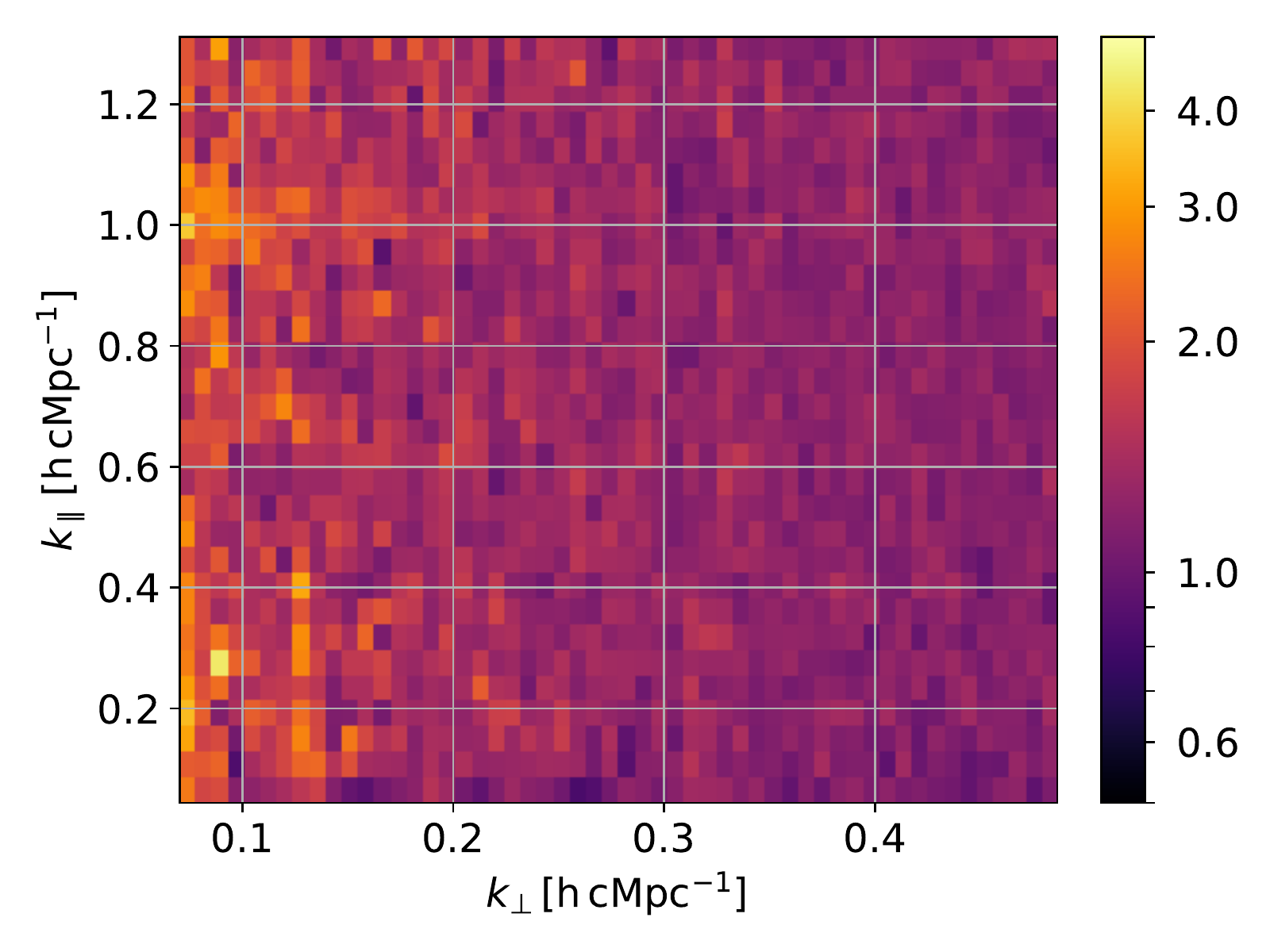}
	\includegraphics[width=0.45\textwidth]{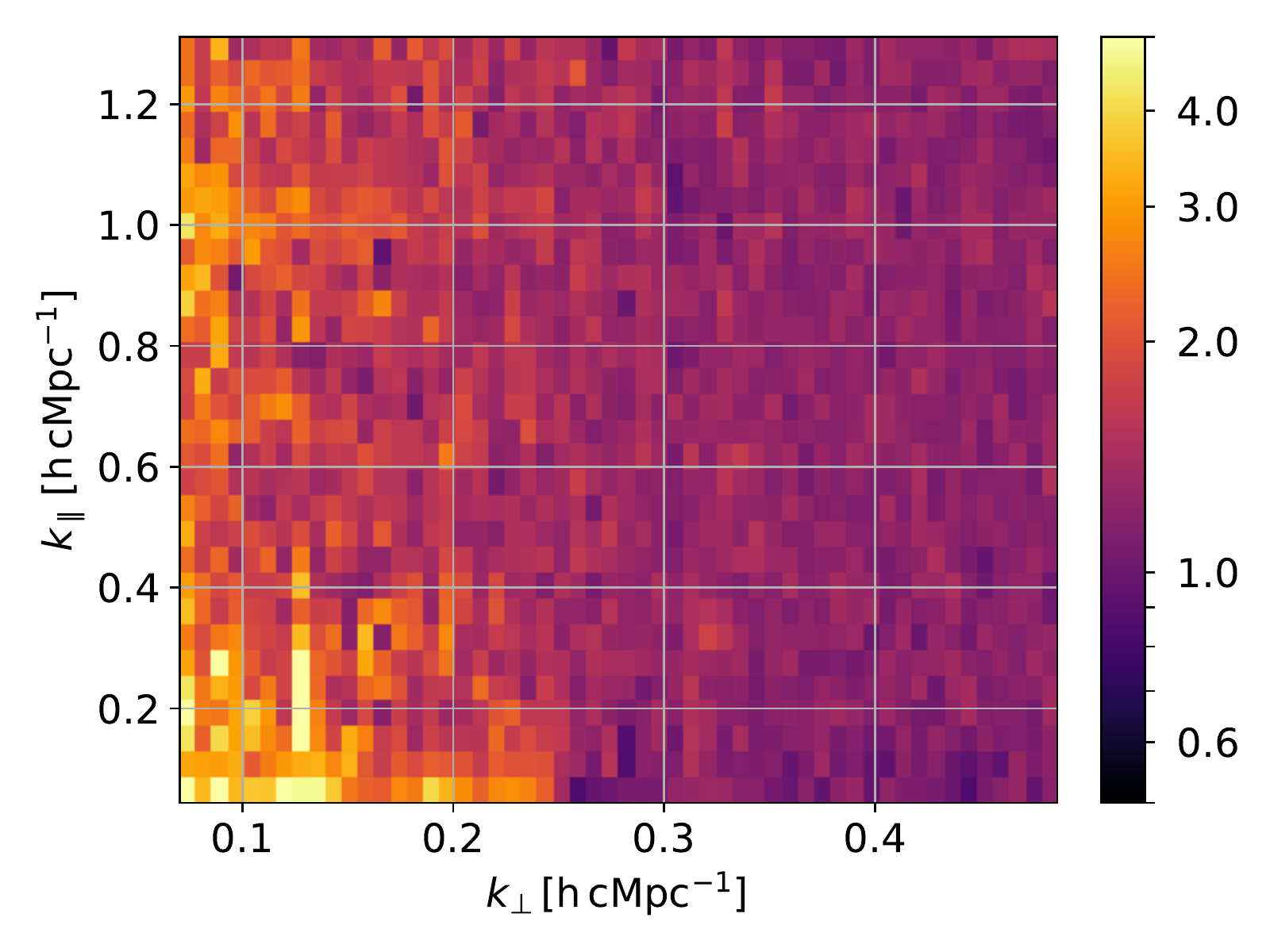}
    \caption{Ratio of the Stokes V power spectra after DD calibration with strong regularisation over the input noise. Left: No baseline cut during calibration. Right: Excluding baselines shorter than $250~\lambda$ from calibration. }
    \label{fig:leverage2}
\end{figure*}

We repeat the tests in Section
\ref{section:simulation}
on the same simulated data. This time, we only use a single realisation of the simulated signal,
corresponding to the brightest selected signal with a relative amplitude of ten. Additionally, we use the limited bandwidth and optimised
regularisation parameters as described in the previous section. As before, we analyse the Stokes-V correlations to understand the effect of calibration on an almost noise-like signal. The Stokes-V power ratio plots with high
regularisation without and with the baseline cut are shown in the left and right panel of Fig.~\ref{fig:leverage2}, respectively. Table \ref{tab:suppression} summarises the
results of the suppression and solver noise measurements for four different
calibration schemes. The schemes have different baseline cuts and levels of
regularisation. The lowest stokes V power spectrum ratio is achieved when no baseline cut is applied, as
expected. However, even with a high gain regularisation level, a considerable
level of signal suppression is observed ($35\%$ suppression), although it is much
lower than without strong regularisation. This remaining signal suppression is likely due to the fact that convergence is too slow for practical purposes, therefore maximum smoothness is not reached within the limited number of fourty iterations.  This is in agreement with the theoretical considerations by \cite{Sardarabadi19}. 
In Figure \ref{fig:solutionsDD_low}, we show the amplitude solutions for all four calibration scenarios of a single time slot, station and direction. The reduction of spectral variations with increased regularisation is visible and even more pronounced in the time averaged delay power spectrum of the complex gains of the same station and direction in the same figure. 
From these results we conclude that combining a baseline cut with a high level of regularisation to enforce smooth gain
solutions is optimal. When using these settings, the signal suppression on the
baselines excluded from calibration is close to zero, as expected. Moreover, the extra noise power on the shorter baselines due to overfitting is reduced by about a factor four with respect to the test using low regularisation. Also, the sharp break that can be observed in Fig.~\ref{fig:leverage1} at the $k_{\perp}$ value corresponding to a baseline length of $250 \lambda$ is hardly visible in the right panel of Fig.~\ref{fig:leverage2}.  We conclude that this way an optimal bias-variance trade off can be reached.

Note that the way our simulated gains were constructed, namely from real data calibration, the input gains were not spectrally smooth. This is shown in figure \ref{fig:solutionsDD_low}, where the input gains in this example show similar structure to the gains of the initial test with low regularisation. True spectral structure in the gains will introduce additional noise power when enforcing maximal smoothness, an effect that needs to be carefully considered when working with real data. It is however expected that the true gains are spectrally smooth scales up to several MHz in real data. In our analysis we are not able to directly detect this additional noise. In principle it has been added to the power spectra in figure \ref{fig:leverage2}, such that our conclusion about the reduction of the overfitting effect using strong regularisation might be an underestimation. In fact, our result strengthens the conclusion that limiting the number of degrees of freedom in calibration is more important to reduce signal suppression and noise enhancement, than reaching the \emph{True} gains. 

\section{Discussion}\label{section:conclusion}
We studied the effect of using DD calibration with a large number of free parameters (i.e. directions and spectral channels) on the noise
power as well as on a simulated 21-cm signal added to a typical LOFAR EoR data set pointing at the NCP. Although some parameters, such as the model components and regularisation per direction were optimised for the chosen field and bandwidth, the main conclusion of our analysis are valid for any LOFAR EoR observation. We investigated the effects of excluding the shorter baselines that are used in the final power spectrum analysis ($< 250\lambda$) from the calibration as well as the level of regularisation of calibration solutions. The regularisation constrains the gain solutions
to be more smooth in frequency. We find that including all the baselines
during calibration, even when using a model of the bright diffuse emission on
those baselines, leads to an unacceptable level of signal suppression. Calibrating without those baselines that will
be used in the final 21-cm signal power spectrum removes signal suppression. However,
excluding short baselines from the calibration also increases the excess noise power
on those same baselines. This is known as the bias-variance trade off. Forcing the solutions to be spectrally smooth, on the other hand, reduces signal suppression significantly, and limits the excess noise. The origin of signal suppression and noise enhancement due to calibration can be the lack of convergence to the \emph{True} gains, the large number of degrees of freedom and sky model incompleteness. For the latter the current analysis gives no definite answer, since most of the simulations were done with a perfect sky model, but it needs to be concerned in real data analysis. We found that convergence to smooth parameters could not be reached in practice on real data, even with a very large number of iterations. The simulate gains in our analysis contained spectral structure, but constraining the solutions to be spectrally smooth did not enhance the noise in our limited data set, strengthening the conclusion that the biggest impact comes from limiting the number of free parameters. Reducing the number of degrees of freedom by enforcing strong regularisation clearly leads to lower signal loss and less noise enhancement. 

We conclude that, for DD
calibration of a typical LOFAR 21-cm data set, it is necessary to use a baseline cut, separating the baselines used for calibration from those used for signal extraction, as well as significantly increase the level of regularisation. We also show that the conservative strategy chosen for the LOFAR EoR analysis by \cite{Mertens2020} does not lead to signal suppression during the calibration step, while the increased noise level due to overfitting is significantly reduced with respect to the strategy presented earlier by \cite{Patil17}. An incomplete sky model may further increase the level of noise due to overfitting or the level of signal suppression if no baseline cut is applied. In the presented study a perfect sky model (apart from the 21-cm signal) was assumed, therefore a lower limit on the level of suppression was established.  Further testing is necessary to determine whether the baseline cut can be safely removed when using an improved model of the sky. However, as shown by the results in this work, the use of an optimised regularisation parameter reduces the solver noise to a lower level (about a factor four in power) and removing the cut is therefore less pressing. 

Besides reducing the parameter
space by enforcing smooth solutions in frequency space, we have additionally shown
that, in real data, the gain parameters never converge to a polynomial
curve for some of the clusters. This is specifically the case for clusters near the first null of the
beam. Including a beam model during DD calibration could therefore improve results further. Since DD effects are expected to be smooth over the field of view, another way to decrease the degrees of freedom in the calibration is to constrain the solutions to be both spectrally and directionally smooth.

The final calibration scheme described here also does not yet
include constrained solutions for the initial two-directional calibration. The
reason for this is that cable reflections and bandpass effects lead to DI gain
variations on short frequency scales. These can not be taken into account with a
third-order Bernstein polynomial. However, as was discussed by
\cite{Barry16} and  \cite{Ewall17}, correcting the data with spurious rapid
varying gain solutions, e.g. due to an incomplete sky model, could introduce
artificial enhanced power on small frequency scales that cannot
be reduced in the subsequent steps of the calibration. Contrary to solver noise, this model incompleteness induced power leakage is correlated and therefore, can not be integrated down by adding observations. Although in DD calibration the final visibilities are not multiplied with the calibration gains, the gain multiplied sky model is subtracted. Therefore, the correlated enhanced power at high k values, arising from incorrect spectral structure in the calibration gains related to model incompleteness, is expected to be less pronounced but could still be present if the DD gains are not forced to be spectrally smooth.  In \cite{Mertens2020} it was shown that with the current settings, a large
fraction of excess noise power (above the theoretical noise power) remains, which needs further investigation (Gan~et~al.~in prep.). In future analyses, we will investigate whether adding extra constraints during our initial calibration or during the DD calibration will further reduce the excess noise power and is able to limit the leakage of power due to model incompleteness. 

We note that, the results in this paper focus on the specific case of data from the LOFAR EoR project. However, the conclusions are also applicable to other experiments that try to 
recover faint signals using various forms of calibration with a large number of free parameters, and to the SKA, that has a design similar to that of LOFAR.

\section*{Acknowledgements}
This result is part of a project that receives funding from the European Research Council (ERC) under the European Union's Horizon 2020 research and innovation programme (Grant agreement No. 884760). EC acknowledges the support of a Royal Society Dorothy Hodgkin Fellowship and a Royal Society Enhancement Award. 
ITI was supported by the Science and Technology Facilities
Council [grant numbers ST/I000976/1 and ST/T000473/1]
and the Southeast Physics Network (SEP-Net). AG would like to acknowledge IUCAA, Pune for providing support through the associateship programme.

\section*{Data Availability}
The data underlying this article will be shared on reasonable request to the corresponding author.




\bibliographystyle{mnras}
\bibliography{mybibliography.bib} 





\bsp	
\label{lastpage}
\end{document}